\documentclass[%
superscriptaddress,
reprint,
showpacs,
 amsmath,amssymb,
 aps,
pra,
]{revtex4-1}

\usepackage{color}
\usepackage{graphicx}
\usepackage{dcolumn}
\usepackage{bm}
\usepackage{amssymb}

\begin{document}

\title{Enhanced spin coherence of rubidium atoms in solid parahydrogen} 

\author{Sunil Upadhyay}
\affiliation{Department of Physics, University of Nevada, Reno NV 89557, USA}
\author{Ugne Dargyte}
\affiliation{Department of Physics, University of Nevada, Reno NV 89557, USA}
\author{Robert P. Prater}
\affiliation{Department of Physics, University of Nevada, Reno NV 89557, USA}
\author{Vsevolod D. Dergachev}
\affiliation{Department of Chemistry, University of Nevada, Reno NV 89557, USA}
\author{Sergey A. Varganov}
\affiliation{Department of Chemistry, University of Nevada, Reno NV 89557, USA}
\author{Timur V. Tscherbul}
\affiliation{Department of Physics, University of Nevada, Reno NV 89557, USA}
\author{David Patterson}
\affiliation{Broida Hall, University of California, Santa Barbara, Santa Barbara, California 93106, USA}
\author{Jonathan D. Weinstein}
\email{weinstein@physics.unr.edu}
\homepage{http://www.physics.unr.edu/xap/}
\affiliation{Department of Physics, University of Nevada, Reno NV 89557, USA}


\begin{abstract}

We measure the transverse relaxation of the spin state of an ensemble of ground-state rubidium atoms trapped in solid parahydrogen at cryogenic temperatures. We find the spin dephasing time of the ensemble (T$_2^*$) is limited by   inhomogeneous broadening. 
We determine that this broadening is dominated by electrostatic interactions with the host matrix, and  can be reduced by preparing nonclassical spin superposition states.
 Driving these superposition states gives significantly narrower electron paramagnetic resonance lines and the longest reported electron spin $T_2^*$  in any solid-phase system other than solid helium. 

\end{abstract}


\maketitle


Measuring the energy splitting between Zeeman levels is at the heart of atomic magnetometry \cite{budker2007optical}, electron paramagntic resonance (EPR) spectroscopy \cite{al1964electron},  and fundamental physics measurements \cite{Andreev2018, pustelny2013global, brown2010new}. 
For an ensemble of $N$ atoms, the shot-noise limited precision of a single measurement is $\sigma_E \sim \frac{\hbar}{T_2^* \sqrt{N}}$  \cite{budker2007optical}, where T$_2^*$ is the ensemble's spin  dephasing time. 
%
%
%
%
In this paper, we show that rubidium atoms in parahydrogen have favorable T$_2^*$ times for a solid state electron spin ensemble. Moreover, their T$_2^*$ can be further extended by using nonclassical superposition states instead of traditional Larmor precession states.


Our apparatus is similar to that described in Refs. \cite{upadhyay2016longitudinal, kanagin2013optical}. We grow our crystal by co-depositing hydrogen and rubidium gases onto a cryogenically-cooled sapphire window at 3 Kelvin.  We enrich the parahydrogen fraction of hydrogen by flowing the gas over a cryogenically-cooled catalyst. In the data presented in this paper, the orthohydrogen fraction is $< 10^{-4}$. Typical thicknesses of the doped crystals are $\sim 0.3$~mm. 
We use natural-isotopic-abundance rubidium; typical rubidium densities are on the order of $10^{17}$~cm$^{-3}$, or a few ppm.

\begin{figure}[ht]
    \begin{center}
    \includegraphics[width=\linewidth]{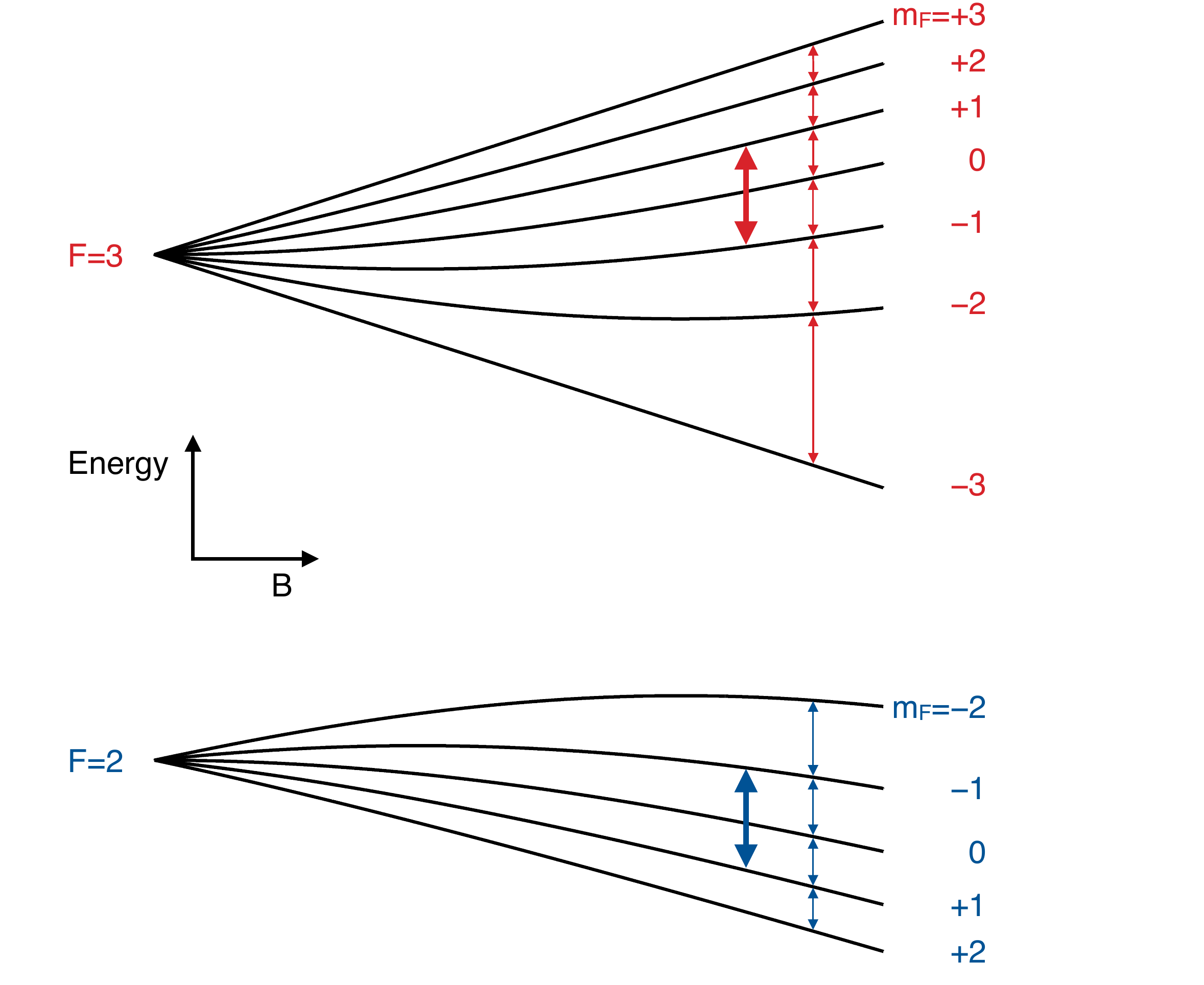}
    \caption{ 
Schematic of the Zeeman levels of gas-phase $^{85}$Rb, showing some of the relevant transitions of Fig. \ref{fig:spectrum}. The energy eigenstates (black) are labeled by their low-field quantum numbers $F$ and $m_F$, and we refer to them throughout the paper by that terminology. The slender  arrows denote the single-photon transitions. The wide arrows denote the two-photon transitions between states which are approximate time reversals of each other; each two-photon transition is shown as a single arrow. 
To better illustrate the nondegenerate frequencies of the transitions, the Zeeman levels are plotted over a larger range of magnetic fields than used in this experiment; likewise the transition arrows are horizontally offset  for ease of viewing.
    \label{fig:HFschematic}}
    \end{center}
\end{figure}

We apply a static ``bias'' magnetic field ($B_z$) normal to the surface of the crystal. We polarize the spin state of the implanted Rb atoms by optically pumping the atoms with a circularly-polarized laser. We measure the polarization through circular dichroism, measuring the relative transmission of left-hand- and right-hand-circularly-polarized light (LHC and RHC). 
We drive transitions between Zeeman states with transverse RF magnetic fields generated by a wire a few~mm above the surface of the crystal.
We take data with bias fields ranging  from 40 to 120 Gauss, giving Zeeman shifts that are small compared to the hyperfine splitting, but sufficiently large that transitions between  different Zeeman levels can be spectrally resolved. The level structure of ground-state $^{85}$Rb is shown in Fig. \ref{fig:HFschematic}.

We  measure rubidium's transverse relaxation time by free-induction-decay (FID) decay measurements. After polarizing the spin through optical pumping, an RF pulse is applied to induce Larmor precession. The Larmor precession and its decay are measured optically via circular dichroism  \cite{budker2007optical}.
The measured values of T$_2^*$ are shorter than our  spin-echo measurements of T$_2$ by over an order of magnitude, indicating that the dominant limit on T$_2^*$ comes from static inhomogneous broadening. Significant inhomogeneous broadening is not surprising, given that our matrix growth conditions are expected to produce polycrystalline parahydogen  \cite{tam1999high, raston2013high}.

For a magnetically-pure host such as parahydrogen, we expect that this inhomogenous broadening  is dominated by electrostatic interactions. 
Due to the symmetry of electrostatic interactions under time reversal, we can find  quantum mechanical superposition states that are more resistant to inhomogenous broadening than Larmor precession states.
We note that the Hamiltonian for electrostatic interactions is unchanged under time reversal. 
Thus, to first order in the perturbation, the electrostatic energy shift of a state $|\psi \rangle$ and its time-reversed state $|\tilde{\psi} \rangle$ must be identical. A superposition state of Zeeman levels which are time-reversals of each other will have a reduced inhomogeneous broadening.
For free-space Rb atoms, $|F, \pm m_F\rangle$ pairs are time reversals of each other in the low-field limit (the stretched states $|F=I+J, m_F = \pm F\rangle$ are time reversals of each other at all fields). We can use multi-photon transitions to prepare superpositions of this kind, as shown in Fig. \ref{fig:HFschematic}.

Superpositions of these states cannot be studied by FID techniques. Only superpositions of Zeeman levels which differ by $\Delta m = 1$ give rise to Larmor precession. For the rubidium superpositions of interest, the expectation value of the spin projection along a transverse axis is zero. Thus there is no \emph{literal} ``transverse spin relaxation'' time, as the ensemble's transverse spin is always zero. However, like any other two-level system, a superposition of $|m_F=+1\rangle$ and $|m_F=-1\rangle$ has a well-defined dephasing time.

To measure T$_2^*$ we use ``depolarization spectroscopy'', wherein we polarize the atoms by optical pumping, then continuously measure the circular dichroism signal as we scan the RF frequency across the resonances. When the frequency is on resonance between two $m_F$ energy eigenstates, population is transferred between them and the polarization signal changes. In the limit of low RF power and a slow frequency sweep, the linewidth of the transition provides a measurement of the inhomogenous broadening and hence T$_2^*$.

\begin{figure}[ht]
    \begin{center}
    \includegraphics[width=\linewidth]{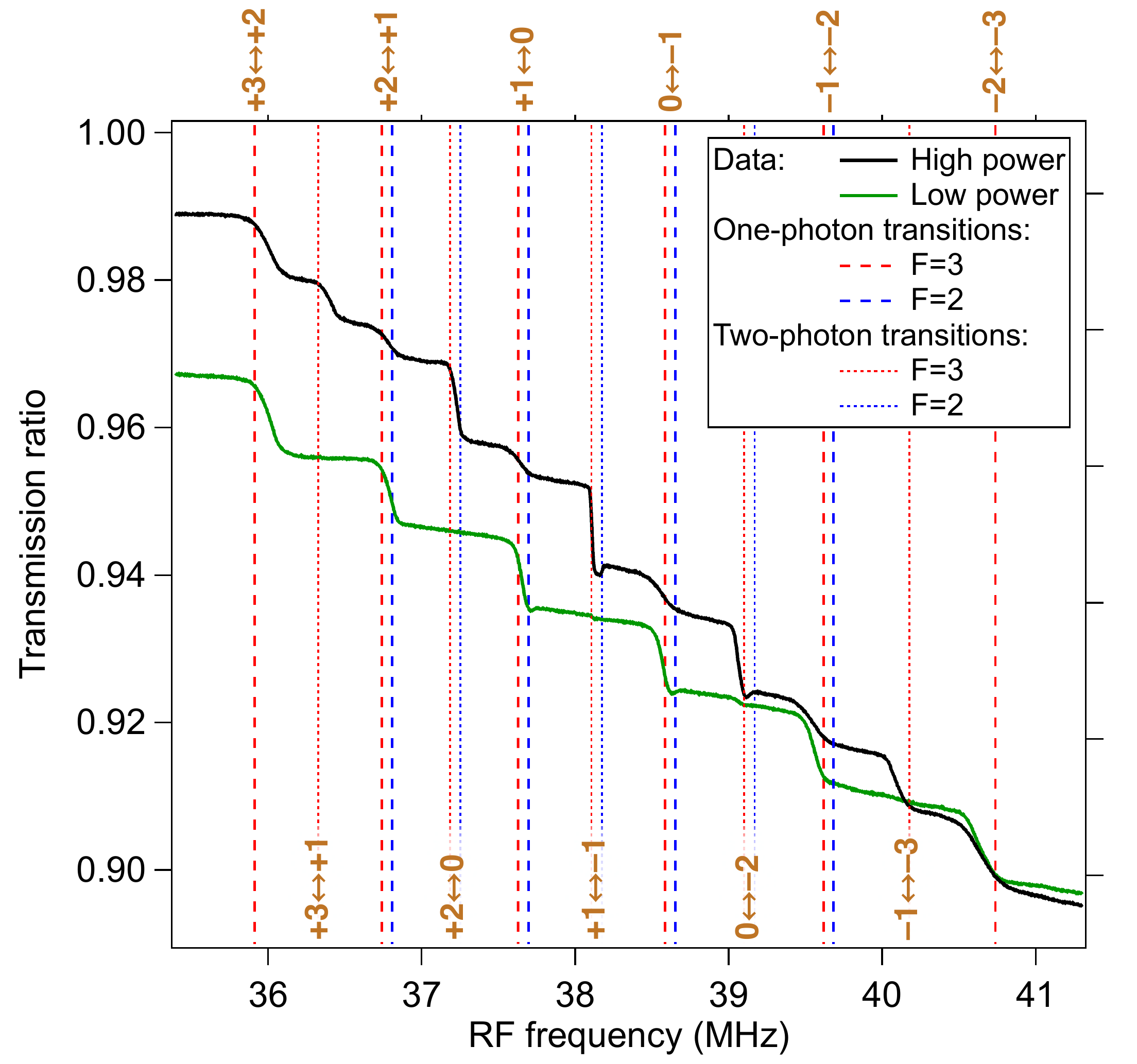}
    \caption{ 
   Depolarization spectroscopy signals, as discussed in the text, taken with $B_z= 82$~G. The vertical axis measures the ratio of the transmission of LHC and RHC probe beams; under conditions of no spin polarization the ratio is 1.
This signal is plotted as a function of the RF frequency; in this data the RF is swept from high to low frequency.
  The dashed vertical lines mark the calculated single-photon and two-photon transition frequencies for gas-phase $^{85}$Rb \cite{steckRb85}; the yellow labels denote $m_F \leftrightarrow m_F'$.
    \label{fig:spectrum}}
    \end{center}
\end{figure}

Fig. \ref{fig:spectrum} shows depolarization data for $^{85}$Rb at  two different RF powers. 
%
For the low-power sweep, we see a change in the polarization signal at each expected single-photon resonance frequency. The broadening is sufficiently large that the $F=2$ transitions are not fully resolved from the $F=3$ transitions.
 At higher powers, the two-photon transitions become observable. 
 The $+1 \leftrightarrow -1$ transitions at the center of the spectrum 
 are significantly narrower than all other one- and two-photon transitions. This is precisely as expected for inhomogenous electrostatic broadening, as they are the only two-photon transitions between time-reversed states. 
 For  $+1 \leftrightarrow -1$, the $F=2$ and $F=3$ transitions are cleanly resolved; the two transitions cause the circular dichroism signal to change in opposite directions.

\begin{figure}[ht]
    \begin{center}
    \includegraphics[width=\linewidth]{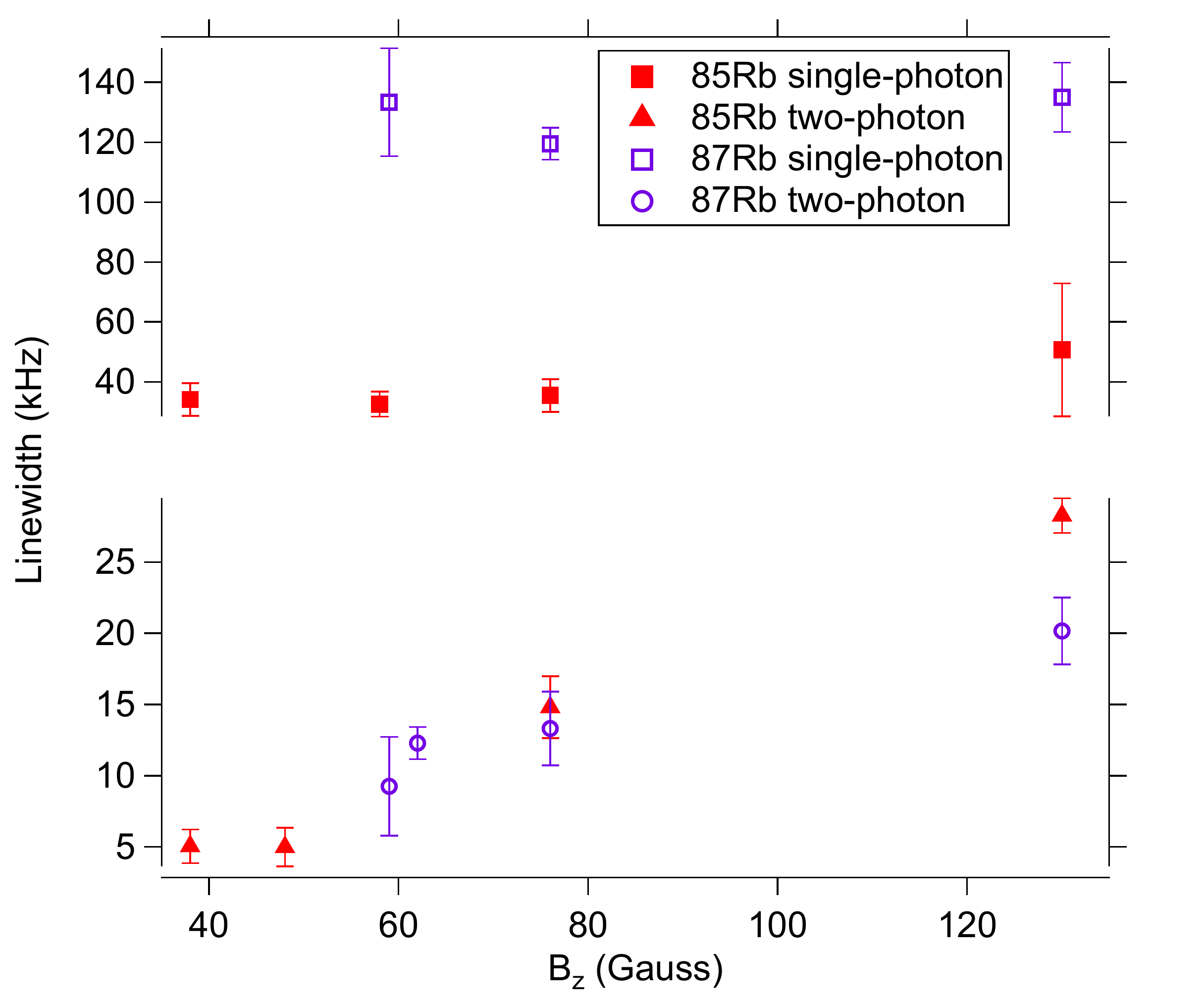}
    \caption{ 
Measured  linewidths of $^{85}$Rb in the $F=3$ hyperfine state and $^{87}$Rb in the $F=2$ hyperfine state.  
For the single-photon linewidth, we plot the average of the $|m_F=\pm 1\rangle \leftrightarrow |m_F=0 \rangle$ transitions. For the two-photon linewidth we plot the linewidth of  the $+1 \leftrightarrow -1$ transition.
    \label{fig:linewidths}}
    \end{center}
\end{figure}

We extract linewidths from this data by assuming the inhomogenous broadening is Gaussian and fitting each transition in the depolarization spectrum to a corresponding error function. The extracted linewidths as a function of magnetic field are shown in Fig. \ref{fig:linewidths} for $^{85}$Rb and $^{87}$Rb. 
These linewidths reflect low-field (i.e. non-power-broadened) values.
For the single photon transitions, the linewidths measured through depolarization spectroscopy match those from FID measurements to within our experimental error.

Examining the single-photon linewidths in Fig. \ref{fig:linewidths}, we see that $^{87}$Rb exhibits more broadening than $^{85}$Rb. This is as one would naively expect for shifts that are electrostatic in nature, as tensor Stark shifts are larger for  ground-state $^{87}$Rb ($F=3$) than for $^{85}$Rb  ($F=2$)\cite{dzuba2010hyperfine}. For a static electric field, calculations predict the $\pm1 \leftrightarrow 0$ transitions in 
$^{87}$Rb ($F=3$) would show 3.2 times the tensor Stark shift of $^{85}$Rb  ($F=2$)\cite{dzuba2010hyperfine}, consistent with the differences seen in the linewidths of Fig. \ref{fig:linewidths}.
Similarly, we find single-photon transitions between states of higher $|m|$ (not shown in Fig. \ref{fig:linewidths}) exhibit more broadening than transitions between states of lower $|m|$; tensor Stark shifts scale as $|m|^2$.

We attribute the  dominant contribution to the linewidth of the $+1 \leftrightarrow -1$ transitions to  residual electrostatic broadening that arises because these energy eigenstates are not perfect time-reversals of each other. While we do not know the exact form of the tensor interaction with the parahydrogen host, we can use perturbation theory to qualitatively predict the linewidth of the $+1 \leftrightarrow -1$ transition. We consider an atom with hyperfine constant of $A$, a Zeeman splitting of $z$, and an interaction with the matrix which is symmetric under time-reversal and on the order of $m$. We consider the limit that $A \gg z \gg m$, and only include perturbations from $m$ to lowest-order in perturbation theory. In this limit, we expect the linewidth of the $+1 \leftrightarrow -1$ transition to be smaller than the single-photon transitions by a factor of $\sim \frac{z}{A}$ (ignoring numerical prefactors). The residual broadening is due to the breakdown of the time-reversal symmetry of the $|+1\rangle$ and $|-1\rangle$ energy eigenstates at nonzero magnetic field.
This model qualitatively agrees with our linewidth measurements (Fig \ref{fig:linewidths}), and explains the observed dependence of the $+1 \leftrightarrow -1$ linewidth on the magnetic field. We note that measurements of $^{85}$Rb as a function of rubidium density indicates that dipolar broadening does not contribute significantly to the linewidth.

The only energy eigenstates of the free atom which are time-reversals of each other at all magnetic fields are the ``stretched states'': $|F=3, m_F = \pm 3 \rangle$ for $^{85}$Rb and $|F=2, m_F = \pm 2 \rangle$ for $^{87}$Rb.  Inhomogenous electrostatic broadening should be further suppressed for these states. We have not been successful in observing the six-photon $+3 \leftrightarrow -3$ transition in $^{85}$Rb; we suspect this is due to insufficient RF power. Observations of the four-photon $+2 \leftrightarrow -2$ transition in $^{87}$Rb at 60 G indicated narrower lines, but not at a statistically significant level. We suspect the measured linewidth of the $^{87}$Rb stretched state transition is limited by technical limitations of our apparatus (magnetic field gradients, as well as magnetic field instabilities which prevent averaging) and our measurement technique.  


To estimate the matrix shifts of the alkali-metal atoms trapped in solid $p$-H$_2$, we use a third-order perturbative expression \cite{Adrian:60} 
 \begin{multline}\label{PT3}
\Delta E_{hf}= \sum_{ij} \sum_{kl} \frac{\langle 00| V_{dd}| ij\rangle \langle ij| H_{hf} |kl\rangle \langle kl| V_{dd} |00\rangle   }{ (E_{00}-E_{ij})(E_{00}-E_{kl}) }  
\\ - \langle 00 | H_{hf}| 00\rangle \sum_{kl} \frac{\langle 00| V_{dd}| kl\rangle \langle kl| V_{dd} |00\rangle  }{(E_{00}-E_{kl})^2 }  
\end{multline}
where the unperturbed  basis functions $|ij\rangle=|i\rangle_A|j\rangle_{\text{H}_2}$ describe the electronic states of  noninteracting atom $A$ and H$_2$, with energies $E_{kl}=E_k^A + E_l^{H_2}$ and $H_{hf}$ is the atomic hyperfine Hamiltonian \cite{Arimondo:77}. The sums in Eq.~(\ref{PT3}) run over all excited electronic, fine, and hyperfine states of $A$ and H$_2$. The first term in Eq. (\ref{PT3}) gives rise to an $m_F$-dependent tensor matrix shift of the atomic hyperfine levels \cite{Moroshkin:08} whereas the second term leads to an $m_F$-independent scalar shift $\Delta E_{hf}^s$, which can be estimated by assuming that $E_{00}-E_{ik}\simeq E_A+E_{H_2}$, where $E_A$ is the average excitation energy of atom $A$ and $E_{H_2}$ is that of  H$_2$ \cite{Adrian:60} . 
Using the closure relation to eliminate the summations over the excited states in  the second term of Eq.~(\ref{PT3})  we obtain  \cite{Adrian:60} 
 \begin{equation}\label{scalar_shift}
 \Delta E_{hf}^s \simeq  - \langle 00 | H_{hf}| 00\rangle \left(  \frac{1}{E_A+E_{H_2}} \right) E_\text{disp}
\end{equation}
where $E_\text{disp}=\langle 00|V_{dd}^2|00\rangle /(E_A+E_{H_2})$  is the dispersion interaction energy of $A$ with H$_2$.

 To estimate the tensor matrix shift, we take into account only the diagonal matrix elements of the hyperfine interaction in Eq. (\ref{PT3}) and assume that  $\langle i| H_{hf}| i\rangle \simeq A_{P}$ is independent of the electronic state $i$ and equal to the hyperfine constant of the lowest excited $^2P_{1/2}$-state of atom $A$. Since the hyperfine constants of alkali-metal atoms decrease rapidly with increasing $i$ \cite{Arimondo:77}, these assumptions provide a conservative upper bound to the magnitude of the tensor shift
 \begin{equation}\label{tensor_shift2}
\Delta E_{hf}^t < A_{P} \sum_{i,j}  \frac{ \langle 00| V_{dd} |ij\rangle \langle ij| V_{dd} |00\rangle   }{ (E_{00}-E_{ij})^2 }  =   0.1 \Delta E_{hf}^s
\end{equation}
where the ratio of the tensor to scalar matrix shifts $\Delta E_{hf}^t/{\Delta E_{hf}^s} \leq {A_{P}}/{A_S} = 0.1$ for Rb \cite{Arimondo:77}. This is consistent with the fact that  the third-order tensor Stark shifts of alkali-metal atoms are suppressed by a factor of $\simeq$100 compared to the scalar shifts \cite{Ulzega:06}.  Note also that the ratio of third-order tensor and scalar polarizabilities of atomic Cs, $\alpha^{(3)}_2/\alpha^{(3)}_0 = 0.03$ \cite{Ulzega:06} is 4 times smaller than the upper bound (\ref{tensor_shift2}). 

\begin{figure}[t]
    \begin{center}
    \includegraphics[width=\linewidth]{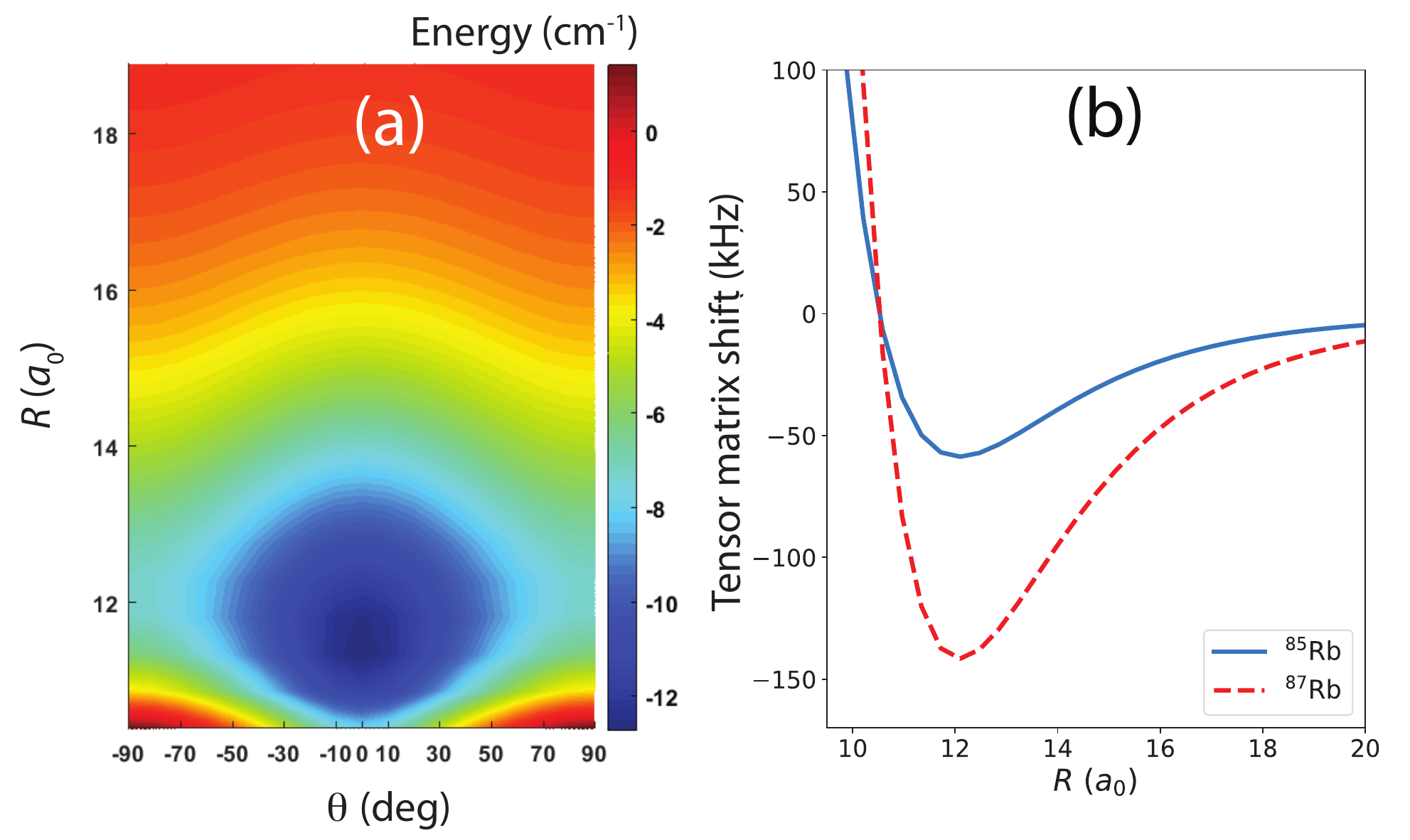}
    \caption{(a) {\it Ab initio} PES for Rb-H$_2$ plotted as a function of the Jacobi coordinates $R$ and $\theta$. (b) Tensor matrix shifts of $^{85}$Rb (full line) and $^{87}$Rb (dashed line) interacting with six impurity $p$-H$_2$ molecules as a function of the atom-impurity distance $R$. } \label{fig:1T}
    \end{center}
\end{figure}

To obtain the dispersion energy needed to estimate the tensor matrix shift via Eq. (\ref{tensor_shift2}), we carried out accurate {\it ab initio} calculations of the Rb-H$_2$ interaction potential using the unrestricted coupled cluster method with single, double and perturbative triple excitations [UCCSD(T)] \cite{Deegan:94}. A large augmented correlation-consistent polarization valence quadruple-$\zeta$ basis set (aug-cc-pVQZ) \cite{Dunning:89} and the ECP28MDF relativistic effective core potential with the [13s10p5d3f] basis set  \cite{Lim:05}  were used for the H and Rb atoms, respectively. The basis set superposition error for Rb-H$_2$ interaction energy was corrected using the standard approach \cite{Boys:70}. All calculations were carried out with the MOLPRO suit of programs \cite{Werner:11} and the PES is expressed in the Jacobi coordinates  $R$ and $\theta$, where $R$ is the distance between the Rb atom and the H$_2$ center of mass, and $\theta$ is the angle between the atom-molecule vector $\mathbf{R}$ and the H$_2$ axis. To obtain the effective Rb-H$_2$ potential $V_0(R)$ used in matrix shifts calculations, we averaged 19 PES cuts corresponding to evenly spaced  values of $\theta\in [0^\text{o},90^\text{o}]$ using the hindered rotor model \cite{Li:10}.  A contour plot of our {\it ab initio} PES shown in Fig.~\ref{fig:1T}(a) demonstrates that the Rb-H$_2$ interaction is weakly anisotropic.

Figure~\ref{fig:1T}(b) shows our calculated upper bounds to the tensor matrix shifts of Rb as a function of the atom-impurity distance calculated for 6 $p$-H$_2$ molecules at a distance $R$ from the impurity. The theoretical bounds are consistent with the  measured values shown in Fig. 3,  reaching their maxima of 59~kHz for $^{85}$Rb  and 142 kHz for  $^{87}$Rb near the minimum of the potential well. The large magnitude of the shift in $^{87}$Rb is due to its  larger hyperfine constant, which exceeds that of $^{85}$Rb by  a factor of 3.4. The  $R$ dependence of the shifts follows that of the Rb-H$_2$ interaction energy, reaching a minimum at $R_e\simeq 12.1\, a_0$ and tending to zero at large $R$. At short values of $R$ to the left of the potential minimum, the dominant mechanism responsible for the matrix shifts is no longer the dispersion interaction, but rather the Pauli exclusion force arising from the overlap of the electronic wavefunctions of Rb and H$_2$ \cite{Adrian:60}. Thus, our matrix shift estimates at $R\ll 12\, a_0$ should not be  considered even qualitatively accurate. 
We also note that the calculated tensor matrix shifts scale with $F$ and $m_F$  as $[3m_F^2-F(F+1)]/[2I(2I+1)]$ due to the second-rank tensor nature of the magnetic dipole hyperfine interaction in the excited atomic states \cite{Ulzega:06}.

In conclusion, we have established that the spin dephasing of Rb atoms in  parahydrogen at densities $\lesssim 10^{18}$~cm$^{-3}$ is dominated by interactions that are electrostatic (or ``T-even'') in nature.  As such, the T$_2^*$ can be significantly increased by replacing traditional Larmor precession states (or traditional single-photon EPR spectroscopy) with superposition states of (or multi-photon transitions between) Zeeman levels that are time-reversals of each other.

This will enable greater resolution in EPR spectroscopy, and is of use for improving ensemble magnetometry \cite{PhysRevX.8.031025, PhysRevLett.113.030803} and for fundamental physics experiments with atoms and molecules in matrices \cite{
weis1997can, vutha2018oriented
}. We note that for these applications, the superposition states we have explored have two advantages: their inhomogenous broadening is reduced and they evolve phase faster than Larmor precession states, limited to a factor of $2F$ for the stretched states. The latter advantage has been explored in  recent work for mechanical oscillators, where larger factors can be achieved \cite{1807.11934}.

Our results are  similar to the ``double quantum coherence magnetometry'' techniques developed for NV centers in diamond \cite{huang2011observation, fang2013high, PhysRevLett.113.030803, PhysRevX.8.031025}, but in a different limit.   NV centers are typically used in the regime where the electrostatic coupling of the spin to the lattice is $\gg$ its coupling to $B_z$; here we work in a limit where the coupling to $B_z$ is $\gg$ the electrostatic coupling to the matrix.
The NV-center limit requires the use of a single-crystal sample with magnetic field parallel to the crystal axis \cite{PhysRevX.8.031025}; in the current work we employ what we expect is a polycrystalline sample \cite{tam1999high, raston2013high}, and see no dependence of the  FID linewidths on the magnetic field direction. We speculate the electrostatic broadening  comes from a combination of inhomogenous trapping sites and inhomogenous crystal axis orientations.

Our narrowest observed linewidth of 5~kHz corresponds to a T$_2^*$ of 60~$\mu$s.  
%
%
We note this ensemble T$_2^*$  is longer than reported values for ensembles of NV centers in diamond \cite{PhysRevX.8.031025}. It is also, surprisingly, an order of magnitude longer than reported for alkali atoms in 
superfluid helium \cite{PhysRevLett.103.035302}. The only condensed-phase electron spin system with longer reported ensemble T$_2^*$ times is atomic cesium in solid He, which was measured at a significantly lower spin density \cite{Weis1996}.

We expect even longer T$_2^*$ times can be obtained in parahydrogen by employing stretch-state superpositions and by producing single-crystal samples through  different growth parameters or sample annealing \cite{
tam1999high, raston2013high}.


\textit{Acknowledgements:}
this material is based upon work supported by the  National Science Foundation under Grant No. PHY--1607072, PHY--1607610, and CHE--1654547.
We gratefully acknowledge assistance in the construction of the experimental apparatus from Wade J. Cline and Carl D. Davidson, Jr..
We thank Andrei Derevianko 
and Amar C. Vutha  for valuable conversations and thank Pierre-Nicholas Roy for providing the hindered rotor code.
\bibliography{ESR2018}

\end{document}